 \documentclass[aps,prl,preprint,groupedaddress]{revtex4-1}
\usepackage{graphicx}
\usepackage{amssymb}
\usepackage{wasysym}

\begin{document}

%\preprint{}

\title{Electron spin polarization by isospin ordering\\in correlated two-layer quantum Hall systems}
\author{L. Tiemann}\email[]{lars.tiemann@phys.ethz.ch}
\affiliation{Solid State Physics Laboratory, ETH Zurich, 8093 Zurich, Switzerland}
\author{M. Hauser}
\affiliation{Max-Planck-Institute for Solid State Research, Heisenbergstr. 1, 70569 Stuttgart, Germany}
\author{W. Wegscheider}
\affiliation{Solid State Physics Laboratory, ETH Zurich, 8093 Zurich, Switzerland}
\date{\today}

\vspace{2cm}
\begin{abstract}
Enhancement of the electron spin polarization in a correlated two-layer two-dimensional electron system at a total Landau level filling factor of one is reported. Using resistively detected nuclear magnetic resonance, we demonstrate that the electron spin polarization of two closely-spaced two-dimensional electron systems becomes maximized when inter-layer Coulomb correlations establish spontaneous isospin ferromagnetic order. This correlation-driven polarization dominates over the spin polarizations of competing single-layer fractional Quantum Hall states under electron density imbalances.
\end{abstract}

\maketitle

The spin is a fundamental quantum mechanical degree of freedom and an intrinsic property of all electrons. Electrons confined to two neighboring but electrically isolated 2D systems (2DESs, or layers) obtain an additional \emph{layer degree of freedom}, characterized by the two-level isospin. In the regime of the quantum Hall effects\cite{vK1980} at large magnetic fields $B$, when the electron densities $n$ of layer $1$ and $2$ are adjusted by electrostatic gating to a total Landau level filling factor of $\nu_{tot} = \nu_1 + \nu_2 = \frac{1}{2}+\frac{1}{2}=1$, with $\nu_i \propto n_i/B$, Coulomb interactions between electrons within each layer become comparable to inter-layer interactions. In an analog to spontaneous easy-plane ferromagnetic ordering, the isospin vector aligns along a particular direction within the easy plane and an extraordinary two-layer state emerges\cite{Halperin1983, Murphy1994} which exhibits spectacular effects. Vanishing Hall and longitudinal resistances\cite{Kellogg2004, Tutuc2004, Yoon2010} and a tunneling characteristic\cite{SpielmanPhD, TiemannNJP2008} reminiscent to the Josephson effect have predominantly been made for $\nu_1 = \nu_2 = \frac{1}{2}$, however, the rich inter-layer physics prevail when $\nu_1 \neq \nu_2$ while $\nu_1 + \nu_2 = 1$.

Here, we report a strong enhancement of the electron spin polarization in a double layer system when inter-layer correlations are induced at $\nu_{tot} = 1$. Using resistively-detected nuclear magnetic resonance (RD-NMR) measurements\cite{Desrat2002, Tiemann2012} we demonstrate that two independent layers that display ordinary fractional quantum Hall (FQH) states will become polarized when we enable inter-layer correlations that establish isospin ferromagnetic order. This is achieved by de-tuning the electron densities from originally $n_1 = n_2$ to ($n_1 \pm\Delta n$) and ($n_2 \mp \Delta n$) so that FQH states emerge. The density imbalance does not destroy the isospin ferromagnetic order, but as electrons swap layers, the isospin vector tilts out of the plane and obtains a finite $z$-component that has the property of an isospin polarization.

The FQH effect \cite{Tsui82, Laughlin83} is a phenomenon governed by Coulomb-mediated correlations among electrons within a single 2DES. It is signaled by a vanishing longitudinal and a quantized Hall resistance at certain filling factor fractions. A double layer will show int\textit{ra}-layer FQH effects but for small separations can also exhibit additional correlation effects originating between electrons with different isospin index. These inter-layer states are described by different physics. The state at $\nu_{tot}$ = 1 is theoretically established for a zero distance limit $(d/\lambda)\longrightarrow 0$ (with layer separation $d$, magnetic length $\lambda = (\hbar/ |e|B)^{\frac{1}{2}}$, elementary charge $e$, Planck's quantum $\hbar$). It assumes fully polarized electron spins to maximize the electron-electron separation and a homogeneous isospin polarization \cite{Halperin1983, Fertig1990, MacDonald1990, Moon1995, Ezawa2013}. Experimental studies have investigated the relationship between electron spin and relative layer separation $d/\lambda$ and found that the critical $(d/\lambda)_c$ where inter-layer correlation effects begin to become relevant will rise when the Zeeman energy is increased\cite{Giudici2010} and that the spin polarization of two correlated layers at $\nu_{tot}=1$ for $d/\lambda \leq (d/\lambda)_c$ is distinctively larger than that of two uncorrelated layers for $d/\lambda > (d/\lambda)_c$ \cite{Kumada2005, Finck2010}. The role of the inter-layer Coulomb interaction can therefore been seen as that of a spin polarizer. By using a combination of density imbalances\cite{Spielman2004, Wiersma2004, Champagne2008, Takase2011, Ding2013} and RD-NMR we demonstrate this polarizing effect experimentally.

Our sample is a standard Hall bar, patterned from a modulation doped double quantum well structure, consisting of two 19 nm wide GaAs quantum wells, separated by an AlAs/GaAs superlattice barrier of 8 nm. The intrinsic electron densities/mobilities are $n_1\approx$ 4.2$\times 10^{10}$~cm$^{-2}$ with $\mu_1\approx$ 850000 cm$^2$/Vs and $n_2\approx$ 5.1$\times 10^{10}$~cm$^{-2}$ with $\mu_2\approx$ 615000 cm$^2$/Vs. The single electron tunnel splitting $\Delta_{SAS}$ is $\approx$900 $\mu$K. We independently study two electrically isolated layers by using depletion gates above or beneath each contact arm near each Ohmic contact\cite{Eisenstein1990b,Rubel1997}. All experiments were performed with a standard lock-in technique by passing oppositely-directed low frequency currents of 1 nA through the two layers (''counter-flow''\cite{Kellogg2004, Tutuc2004, Yoon2010}). This geometry guarantees finite longitudinal resistance, $R_{xx}$, at all ($\nu_1,\nu_2$) to permit resistively-detected NMR. The electron densities in both layers are independently controlled via a top and back gate. The sample is surrounded by a NMR-coil, connected to a radio-frequency (RF) generator. In order to obtain a sufficient NMR signal, experiments had to be performed at 3.53 T near $(d/\lambda)_c$ but where the $\nu_{tot} = 1$ state is still strong.

 \begin{figure}[!ht]
 \includegraphics[width=0.65\textwidth]{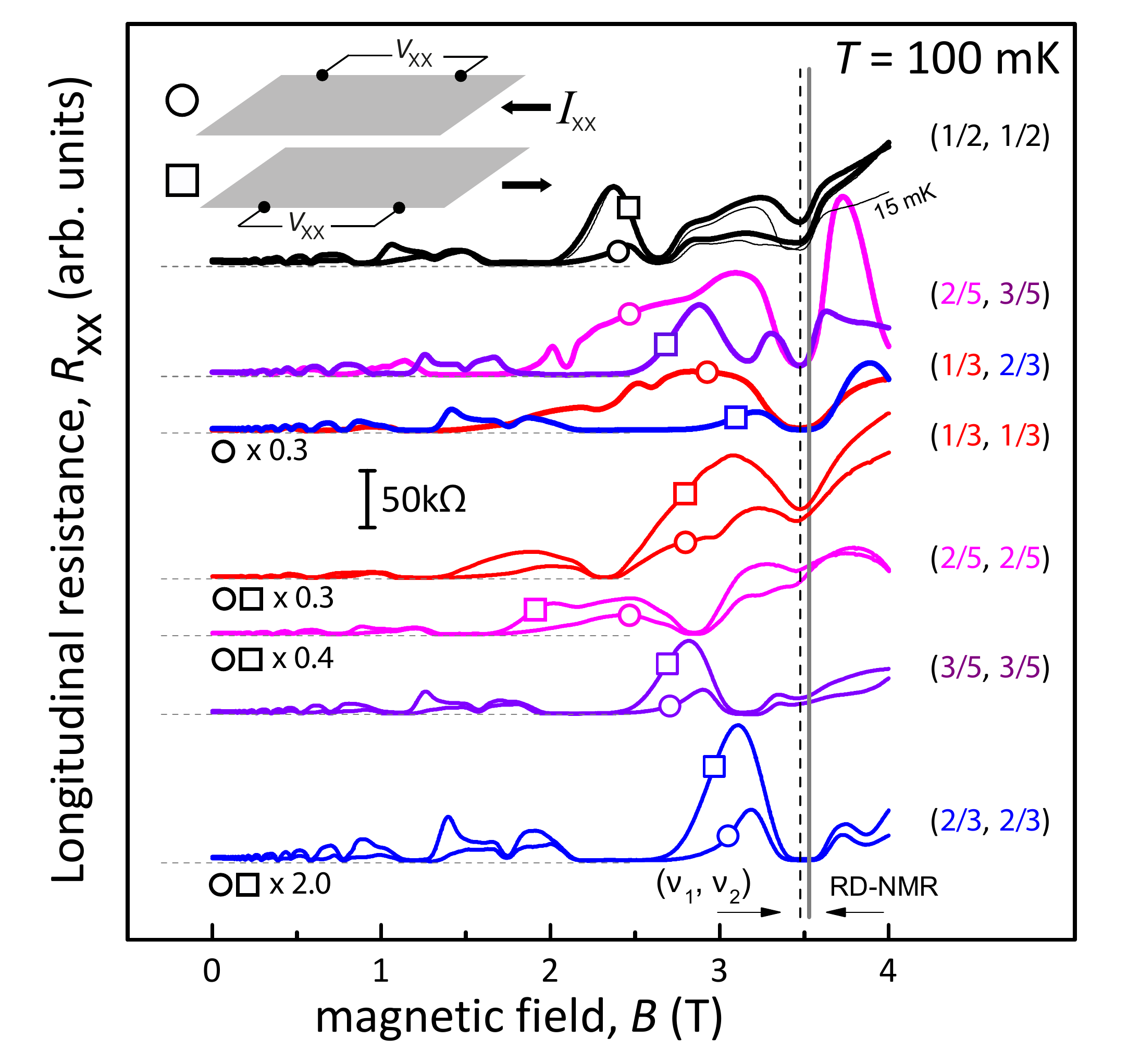}
 \caption{$R_{xx}$ measured independently (inset) in the upper layer ($\Circle$) and lower layer ($\square$) at 100 mK for various $n_1$ and $n_2$ with $I_{xx}$=1 nA (downsweeps with 50 mT/min). The curves have been shifted and scaled for clarity. The thin black line shows transport at 15 mK only for $(\nu_1, \nu_2) =  (\frac{1}{2}, \frac{1}{2})$. The dashed line at 3.48 T indicates $B$ for the filling factors $(\nu_1, \nu_2)$. The solid line at 3.53 T indicates where RD-NMR is performed. For the topmost trace a weak minimum has formed at 3.48 T. De-tuning the densities to ($n_1 - \Delta n$) and ($n_2 +\Delta n$) results in deeper minima. For reference we show uncorrelated combinations with $\nu_1=\nu_2 = \{\frac{1}{3}, \frac{2}{5}, \frac{3}{5}, \frac{2}{3}\}$. For $(\nu_1, \nu_2) = (\frac{2}{5},\frac{2}{5})$ and $(\frac{3}{5},\frac{3}{5})$ FQH minima are almost absent. The magneto-transport traces for $(n_1 + \Delta n)$ and $(n_2 - \Delta n)$ exhibit the same qualitative behaviour and are not shown. Note that for the curves marked $(\frac{1}{3},\frac{1}{3})$ and $(\frac{2}{5},\frac{2}{5})$ another $\nu_{tot}=1$ state appears at lower $B$. \label{fig1}}
 \end{figure}

Figure \ref{fig1} shows $R_{xx}$, simultaneously and independently measured on the two electrically isolated layers for selected values of carefully matched and imbalanced electron densities\footnote{Each layer acts as a gate for the adjacent layer due to ''compressibility'' [Phys. Rev. B 50, 1760 (1994)]. When we create a density imbalance $(n_1 \pm \Delta_n)$ and $(n_2 \mp \Delta n)$, this weak mutual gate effect also changes. For the maximal imbalance, the density would deviate by a few percent from the original (matched) value if we used the same gate voltages. To ensure that we operate at the same $\nu$ under imbalance, we have compensated this effect by carefully adjusting top and back gate voltages}. Magneto-transport was measured while applying the same -12dBm RF power that was used for RD-NMR, resulting in a temperature of 100 mK. For the top-most pair, the densities were set to $n_1 = n_2 \approx 4.3 \times 10^{10}$ cm$^{-2}$, so that at $B$ = 3.48 T a resistance minimum signals the inter-layer $\nu_{tot}$ = 1 state for $(\nu_1, \nu_2) = (\frac{1}{2},\frac{1}{2})$. The $\nu_{tot}$ = 1 state is very temperature-robust\cite{Kellogg2004}, so that the weak resistance minimum we observe is mostly due to the large effective layer separation $d/\lambda \approx$ 2.0. For the subsequent two pairs the electron densities were de-tuned to $(n_1 - \Delta n_i)$ and $(n_2 + \Delta n_i)$. Under this imbalance, $\nu_1$ decreases whereas $\nu_2$ increases and FQH states emerge at certain fractional values while a finite isospin polarization arises. The resulting deepening of the resistance minimum however must not be attributed to a simple coincidence of intra-layer FQH states that emerge at ($\frac{1}{3},\frac{1}{3})$ and ($\frac{2}{3},\frac{2}{3})$. 

The four pairs of curves at the bottom were measured at $n_1 = n_2$ but with higher and lower densities at which inter-layer correlations have given way to ordinary intra-layer QH effects. For  $(\nu_1, \nu_2) = (\frac{2}{5},\frac{2}{5})$  and $(\frac{3}{5},\frac{3}{5})$, we do not see discernable resistance minima which would indicate a FQH state. The deep minimum formed for their combination $(\frac{2}{5},\frac{3}{5})$ however is a direct manifestation of dominant inter-layer physics. The following RD-NMR experiments will demonstrate that also for $(\frac{1}{3},\frac{2}{3})$, the minima at 3.48 T do not represent individual FQH ground states and their distinct spin polarizations but the $\nu_{tot}$ = 1 ground state.

\begin{figure}[!ht]
 \includegraphics[width=0.65\textwidth]{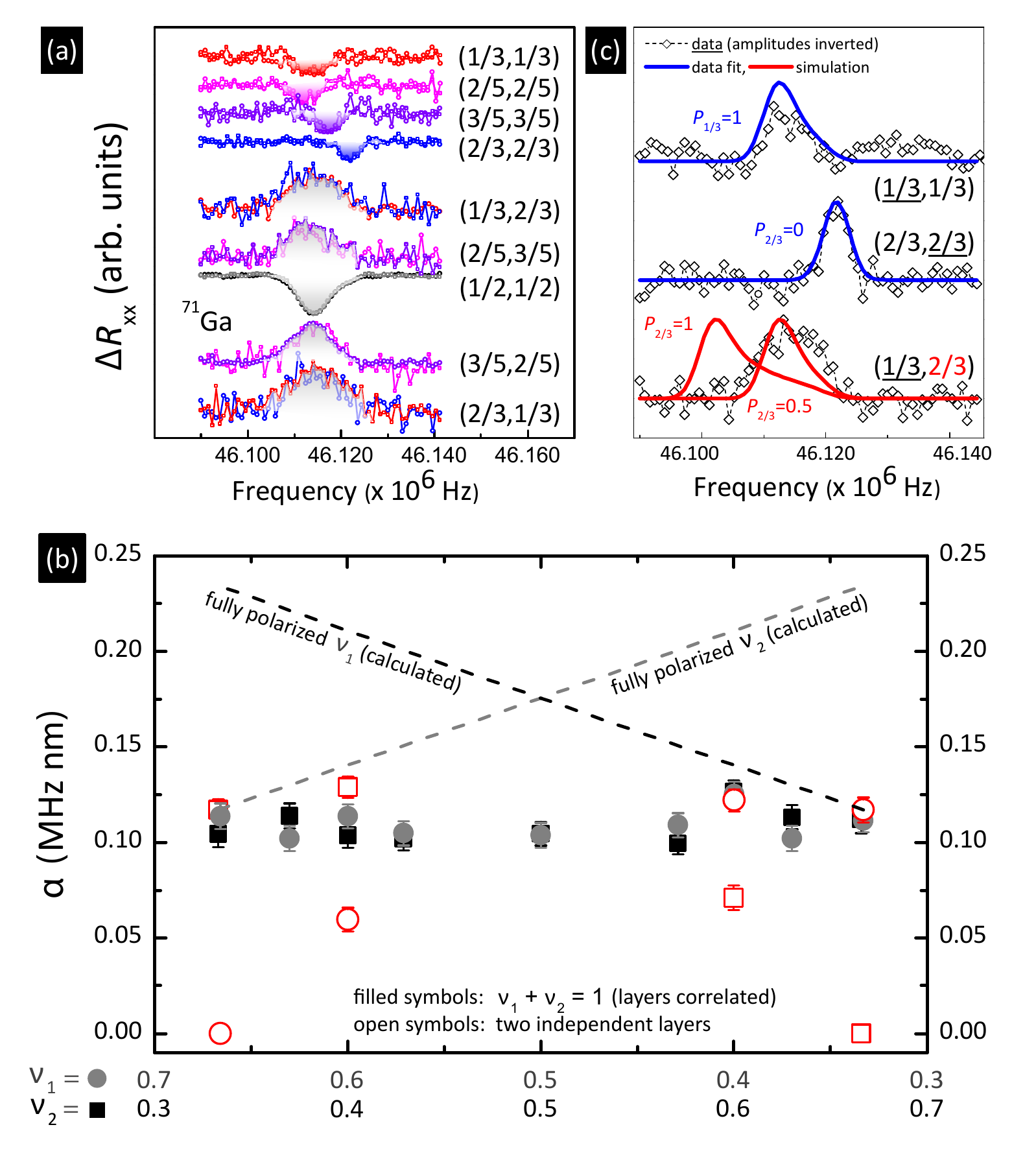}
 \caption{(a) Pairs of RD-NMR spectra for the $^{71}$Ga nuclei measured simultaneously in both layers for different $(\nu_1, \nu_2)$ at 100 mK. The absolute change $\Delta R_{xx}$ (=amplitude) ranges from several tens to several hundred Ohms, depending on $\nu$ and layer index. Even for $\nu_1=\nu_2$, the amplitudes may differ significantly between the layers as a result of different slopes in $R_{xx}$ at $\nu$. To allow visual comparison, the amplitudes were normalized, resulting in different noise levels. Several spectra were averaged or passed through a percentile filter to reduce noise. Transparent shading was added to underline the NMR signal. (b) Spectra were fitted\cite{Tiemann2012} to obtain the Knight shift, expressed as parameter $\alpha$ for the lower layer at $\nu_2$ ($\square$-symbols) and upper layer at $\nu_1$ ($\Circle$). Dashed lines indicate $\alpha$ expected for single fully polarized (uncorrelated) 2DES at $\nu$. (c) Spectra measured at $(\frac{1}{3},\frac{1}{3})$ (fully polarized) and $(\frac{2}{3},\frac{2}{3})$ (unpolarized) are compared to simulated spectra for a half-polarized and fully polarized $\nu=\frac{2}{3}$ system (to maintain clarity, only significant spectra are shown). \label{fig2}}
\end{figure}

The electron spin polarization is encoded in the nuclear resonance frequency, which we access via the hyperfine interaction\cite{LiBook} that couples the magnetic moments of the gallium and arsenide nuclei in the quantum wells to the spin moments of the electrons confined therein. When the 2DES is polarized this coupling reduces the nuclear level splitting for $B>0$ so that resonant transitions between levels occur at a lower frequency. We measure the nuclear resonances by monitoring $R_{xx}$ under RF irradiation which changes in response to the resonant absorption, also through the hyperfine interaction. Nuclear resonances can therefore be detected resistively and contain the electron spin polarization. Figure \ref{fig2}(a) shows pairs of RD-NMR spectra; each spectrum reflects the resonance frequency of the $^{71}$Ga isotopes in one quantum well, with a finite line width that originates from the spatially varying overlap between the nuclei and the electronic wave function \cite{Tiemann2012}. Each pair was measured simultaneously in both layers at 3.53 T for various ($\nu_1$, $\nu_2$) to arise at 3.48 T, hence we actually probe the system at $(\nu_1-\delta\nu_1, \nu_2-\delta\nu_2)$, with $\delta\nu < 0.0095$ for all $\nu$. Spectra were measured at the high field flank of all $R_{xx}$ minima by stepping across the resonance frequency $\omega_0$ of the nuclei with a resolution of $\Delta\omega$ = 0.8 - 1 kHz, i.e., $\omega_0 - N\cdot\Delta\omega < \omega < \omega_0 + N\cdot\Delta\omega$, with $N$ as an integer. Each pulse has a duration of 5-10 s. The nuclei can relax back to thermal equilibrium over $>$ 60 s at a fixed off-resonance frequency between pulses. This method does not rely on dynamical nuclear polarization and is similar to the one used in Refs. \cite{Stern2012, Friess2014}, here, however, $(\nu_1, \nu_2)$ is also used for resistive read-out. For broad minima at all $\nu=\frac{2}{3}$, $\nu_{read}= 0.651$ was used \cite{Tiemann2012}. A constant -12 dBm RF power had to be applied to obtain a reasonable signal at all $\nu$. We will illustrate the main results of our experiment by first focusing on filling factors $\frac{1}{3}$ and $\frac{2}{3}$.

For $(\nu_1, \nu_2)=(\frac{1}{3},\frac{1}{3})$ and $(\nu_1, \nu_2)=(\frac{2}{3},\frac{2}{3})$, the two layers show individual intra-layer FQH states but not the inter-layer $\nu_{tot}$ = 1 state at 3.48 T, and we observe that their respective NMR spectra (first and fourth row in Fig.\ref{fig2}(a)) are shifted with respect to one another. This Knight shift, $K_s$, is the result of distinct electron spin polarizations and can be expressed as $K_s(\nu,z) = \alpha_\nu \cdot |\Psi_\nu(z)|^2$, with parameter $\alpha_\nu = c\cdot \widetilde{n}_{P} \cdot P$, where $c$ is a constant, $P$ the polarization, $\widetilde{n}_{P}$ the effective single layer electron density that contributes to $P$ and $\Psi_\nu(z)$ the subband wave function. The $\frac{1}{3}$ FQH state is always fully polarized, whereas the $\frac{2}{3}$ FQH ground state polarization depends on the ratio of Coulomb to Zeeman energy \cite{Freytag2001}. A narrow 19 nm quantum well at 3.5 T is very likely unpolarized \cite{Yusa2013} but it can undergo a spin transition at large $B$ (Ref. \cite{Eisenstein1990b, Chakraborty1990}), which depends on the finite thickness of the 2DES via the Coulomb interaction \cite{HoeppelPhD, DasSarma1986}. Now we de-tune the densities to obtain $(\frac{1}{3},\frac{2}{3})$ so that strong inter-layer correlations emerge. We find that layer 2 now exhibits a much larger Knight shift as compared to $(\frac{2}{3},\frac{2}{3})$. Increases in the Knight shift are also observed for combinations involving $\frac{2}{5}$ and $\frac{3}{5}$ that add up to $\nu_1 + \nu_2 = 1$. This is an impressive demonstration of how inter-layer correlation physics can polarize formerly unpolarized and uncorrelated layers\footnote{We have scrutinized the temperature dependence, but due to the small Knight shift we found no significant/discernable change in $K_s$ for $(\nu_1,\nu_2) = (\frac{1}{3},\frac{2}{3})$ up to ca. 300 mK at which the NMR signal became too small.}.

We also observe the sign of the RD-NMR amplitude to invert under imbalance. A negative sign results from a larger $R_{xx}$ after resonance whereas a positive sign indicates that the $\nu_{tot} = 1$ state strengthens and broadens when nuclei are depolarized. This observation seems to be in agreement with Refs. \cite{Spielman2005, TracyPhD}, which found that the $\nu_{tot} = 1$ zero bias differential tunneling conductance is strongly enhanced for depolarized nuclei. Close to the phase boundary $(d/\lambda)_c$, the tunneling conductance was reported to be a non-monotonic function of the imbalance \cite{Champagne2008}. Although $\alpha$ appears to be constant under imbalance, the amplitude inversion might be related to the non-monotonic behavior found in tunneling. We also emphasize that without imbalance the state as seen in transport is weak and may not homogenously exist across the Hall bar. For reference we have measured two spectra at the 3.43 T low field flank and found the same qualitative behavior.

We have fitted the spectra to obtain $\alpha_\nu$, a value that represents the Knight shift \cite{Tiemann2012}, $K_s(\nu)$. Figure \ref{fig2}(b) shows $\alpha$ deduced from the spectra shown in Fig.\ref{fig2}(a), including additional intermediate values. When each layer displays an individual FQH state such as $(\nu_1, \nu_2)=\{(\frac{2}{3},\frac{2}{3}), (\frac{2}{5},\frac{2}{5}), (\frac{3}{5},\frac{3}{5})\}$, $\alpha$ varies considerably (open symbols). Filling factor combinations $(\nu_1, \nu_2)$ that enable strong inter-layer correlations with $\nu_1 + \nu_2 = 1$ however show a considerable increase of $\alpha$ towards a maximized polarization (closed symbols). This effect is particularly strong for those combinations with one layer at $\frac{2}{3}$.

To understand the extent of polarization, Fig.\ref{fig2}(c) exemplifies simulated spectra for a $\nu=\frac{2}{3}$ state at various polarizations, $P$. From the electron density, $n_{2/3}$, and $\Psi_{2/3}$, we have calculated $K_s$ and hence can construct the expected spectra for $P_{2/3} = \{0, \frac{1}{2}, 1\}$ \cite{Tiemann2012}. For $P_{2/3} = \frac{1}{2}$, the simulated spectrum coincides with the measured spectrum of $\nu=\frac{1}{3}$ because $K_s\propto n \cdot P$ is the same due to $n_{2/3} = 2\cdot n_{1/3}$. By assuming $P = 1$ for all investigated $\nu$, we can obtain $\alpha$ for a fully polarized and uncorrelated single layer, indicated by the two dashed lines in Fig.\ref{fig2}(b).

Based on our data and simulations we conclude that the $\nu_{tot}=1$ ground state polarization is not determined by the electron spin polarizations of the two individual layers which may display competing intra-layer FQH states. Density imbalances do not influence the electron spin polarization or destroy the isospin order but tilt the isospin vector out of the easy plane where it obtains a finite $z$-component. The two spectra for each imbalanced $(\nu_1, \nu_2)$ are identical in both shape and resulting $\alpha$. This implies that during the transition from an uncorrelated to a correlated system, a certain number of electrons will flip their spin so that the density of spin polarized electrons is constant. These conclusions are in line with the established model of indistinguishable layers at $\nu_{tot} = 1$. We rule out simple single particle tunneling between two quantum Hall systems as a trivial explanation because for $(\frac{1}{3}, \frac{2}{3})$, the resulting spectra would be broader and display two very distinct peaks at the positions of their uncorrelated resonance frequencies. Theory predicts a fully polarized $\nu_{tot}=1$ ground state for $(d/\lambda)\longrightarrow 0$, a limit which never reflects experimental reality. Factors that will influence $\alpha$ are the increased temperature, the larger $d/\lambda$ and the small $\delta\nu$. However, at $\nu_{tot}=1$ the two-layered system is described by a single wave function, so it is unlikely that one layer would display a larger or smaller polarization than the other one as if they were independent.

\begin{figure}[!ht]
\includegraphics[width=0.6\textwidth]{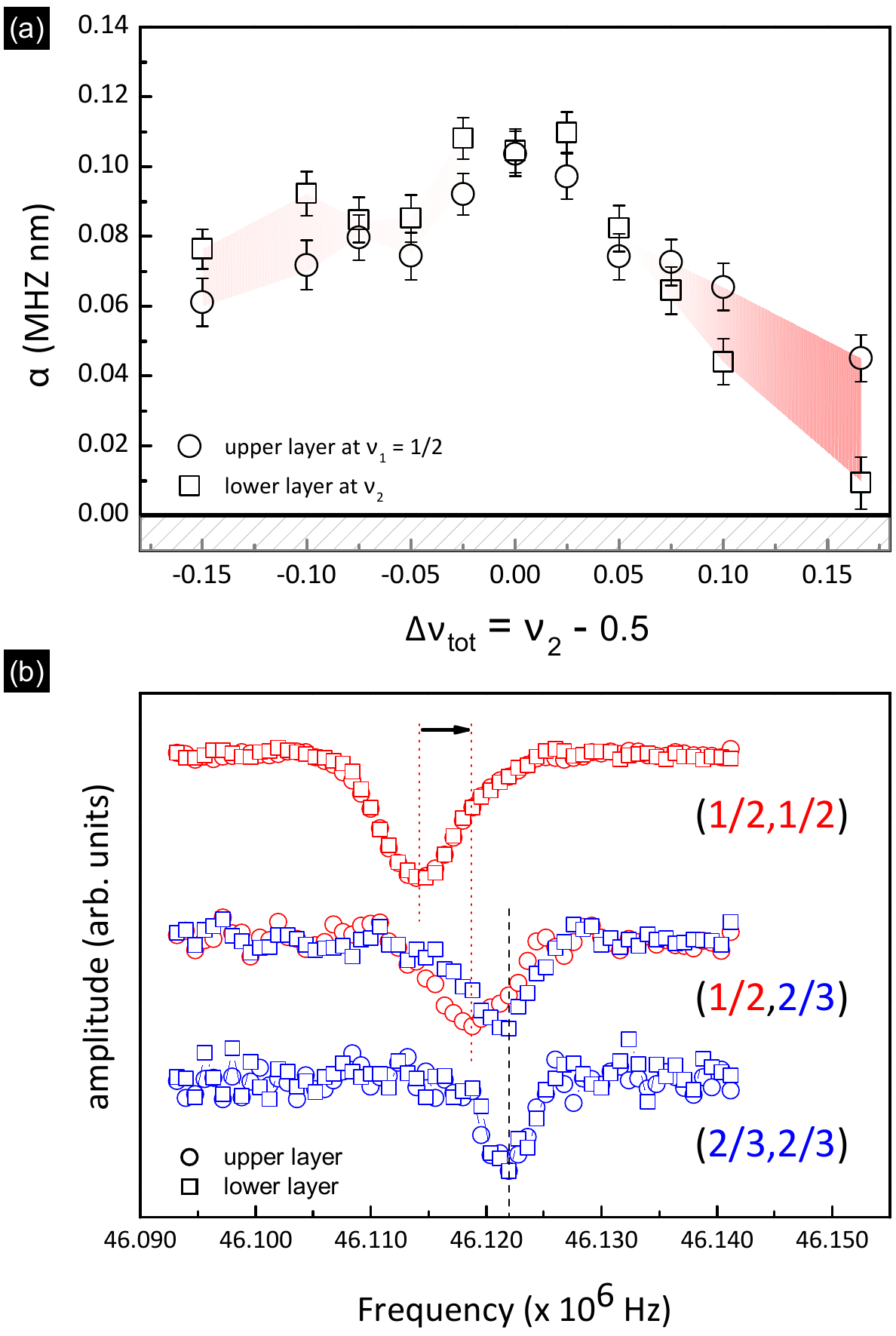}
\caption{(a) $\alpha$ versus $\Delta\nu_{tot} = \nu_2-\frac{1}{2}$, with the upper layer ($\Circle$) at constant $\nu_1=\frac{1}{2}$ and the lower layer ($\square$)
 between $\frac{1}{3} < \nu_2 \leq \frac{2}{3}$ at 100 mK. The color gradient emphasizes the separation of $\alpha$ when the two layers break up into individual quantum Hall systems with distinct spin polarizations. (b) Selected spectra for $(\nu_1, \nu_2) = \{(\frac{1}{2},\frac{1}{2}), (\frac{1}{2},\frac{2}{3}), (\frac{2}{3},\frac{2}{3})\}$, exemplifying this divergence of the spectra for $\nu_{tot} \neq 1$. \label{fig3}}
\end{figure}

The following experiment demonstrates that our NMR measurements indeed probe the $\nu_{tot}=1$ state and that at even $(\frac{1}{2},\frac{1}{2})$ the polarization is increased via the inter-layer correlation. For the data shown in Fig. \ref{fig3}, the upper layer was kept at constant $\nu_1 = \frac{1}{2}$, while the lower layer filling factor was varied between $\frac{1}{3} < \nu_2 \leq \frac{2}{3}$. This measurement is the RD-NMR compatible counterpart to measuring a $d/\lambda_B$ dependence because we rely on a constant $B$. In Fig. \ref{fig3}(a) we plot $\alpha$ versus $\Delta\nu_{tot} = \nu_2-\frac{1}{2}$. For small $\Delta\nu_{tot}$, $\alpha$ for both layers decreases simultaneously and begins to diverge, eventually showing the same quantitative behavior as expected for a single 2DES. For reference, Fig. \ref{fig3}(b) shows selected spectra of $(\nu_1,\nu_2) = \{(\frac{1}{2},\frac{1}{2}), (\frac{1}{2},\frac{2}{3}), (\frac{2}{3},\frac{2}{3})\}$. An individual layer at $\nu =\frac{1}{2}$ is a compressible sea of composite Fermions. Its polarization depends only on the Zeeman energy so that at low $B$ it will be partially polarized\cite{Igor1999}. This behavior agrees with $\alpha$ of layer 1 for $\Delta\nu_{tot}\longrightarrow \pm 0.167$, although there remains a difference in its polarization on $\nu_2$. The lower layer polarization also follows the expected behavior, becoming unpolarized for $\nu_2 \longrightarrow \frac{2}{3}$ ($\Delta\nu_{tot}\longrightarrow +0.167$) and having a maximized polarization for $\nu_2 \longrightarrow \frac{1}{3}$ ($\Delta\nu_{tot}\longrightarrow -0.167)$. By changing $\nu_{tot}$ instead of $d/\lambda_B$ we not only quickly destroy the isospin order but we may also introduce Skyrmions. These transverse fluctuations in the spin field would reduce the measured spin polarization as well. The asymmetric drop of $\alpha$ for larger $\Delta\nu_{tot}$ however suggests that the depolarization is mainly caused by the destruction of the isospin order.

In summary, we have performed RD-NMR measurements in a double quantum well system when both layers are correlated to form the $\nu_{tot}=1$ state. We observe a maximized spin polarization of the correlated system which is indifferent to electron density imbalances and even prevails over unpolarized fractional quantum Hall states that coincide with $\nu_{tot} = 1$. In a polarized system, the electron-electron separation will be maximized via the Pauli-Exclusion Principle. For maximally spaced electrons, the interaction energy will become minimized and electrons can condense into the $\nu_{tot}=1$ ground state.

\begin{acknowledgments}
We would like to acknowledge Werner Dietsche, Timo Hyart and Koji Muraki for discussions and valuable comments on the experiments and this manuscript.
\end{acknowledgments}

%\bibliographystyle{aipauth4-1}
%\bibliography{citations}{}
%\bibliographystyle{apsrev4-1}
\bibliographystyle{apsrev4-1} % Tell bibtex which bibliography style to use
\bibliography{citations_revised}{}

\begin{thebibliography}{39}%
\makeatletter
\providecommand \@ifxundefined [1]{%
 \@ifx{#1\undefined}
}%
\providecommand \@ifnum [1]{%
 \ifnum #1\expandafter \@firstoftwo
 \else \expandafter \@secondoftwo
 \fi
}%
\providecommand \@ifx [1]{%
 \ifx #1\expandafter \@firstoftwo
 \else \expandafter \@secondoftwo
 \fi
}%
\providecommand \natexlab [1]{#1}%
\providecommand \enquote  [1]{``#1''}%
\providecommand \bibnamefont  [1]{#1}%
\providecommand \bibfnamefont [1]{#1}%
\providecommand \citenamefont [1]{#1}%
\providecommand \href@noop [0]{\@secondoftwo}%
\providecommand \href [0]{\begingroup \@sanitize@url \@href}%
\providecommand \@href[1]{\@@startlink{#1}\@@href}%
\providecommand \@@href[1]{\endgroup#1\@@endlink}%
\providecommand \@sanitize@url [0]{\catcode `\\12\catcode `\$12\catcode
  `\&12\catcode `\#12\catcode `\^12\catcode `\_12\catcode `\%12\relax}%
\providecommand \@@startlink[1]{}%
\providecommand \@@endlink[0]{}%
\providecommand \url  [0]{\begingroup\@sanitize@url \@url }%
\providecommand \@url [1]{\endgroup\@href {#1}{\urlprefix }}%
\providecommand \urlprefix  [0]{URL }%
\providecommand \Eprint [0]{\href }%
\providecommand \doibase [0]{http://dx.doi.org/}%
\providecommand \selectlanguage [0]{\@gobble}%
\providecommand \bibinfo  [0]{\@secondoftwo}%
\providecommand \bibfield  [0]{\@secondoftwo}%
\providecommand \translation [1]{[#1]}%
\providecommand \BibitemOpen [0]{}%
\providecommand \bibitemStop [0]{}%
\providecommand \bibitemNoStop [0]{.\EOS\space}%
\providecommand \EOS [0]{\spacefactor3000\relax}%
\providecommand \BibitemShut  [1]{\csname bibitem#1\endcsname}%
\let\auto@bib@innerbib\@empty
%</preamble>
\bibitem [{\citenamefont {von Klitzing}\ \emph {et~al.}(1980)\citenamefont {von
  Klitzing}, \citenamefont {Dorda},\ and\ \citenamefont {Pepper}}]{vK1980}%
  \BibitemOpen
  \bibfield  {author} {\bibinfo {author} {\bibfnamefont {K.}~\bibnamefont {von
  Klitzing}}, \bibinfo {author} {\bibfnamefont {G.}~\bibnamefont {Dorda}}, \
  and\ \bibinfo {author} {\bibfnamefont {M.}~\bibnamefont {Pepper}},\ }\href
  {\doibase 10.1103/PhysRevLett.45.494} {\bibfield  {journal} {\bibinfo
  {journal} {Phys. Rev. Lett.}\ }\textbf {\bibinfo {volume} {45}},\ \bibinfo
  {pages} {494} (\bibinfo {year} {1980})}\BibitemShut {NoStop}%
\bibitem [{\citenamefont {Halperin}(1983)}]{Halperin1983}%
  \BibitemOpen
  \bibfield  {author} {\bibinfo {author} {\bibfnamefont {B.~I.}\ \bibnamefont
  {Halperin}},\ }\href@noop {} {\bibfield  {journal} {\bibinfo  {journal}
  {Helv. Phys. Acta}\ }\textbf {\bibinfo {volume} {56}},\ \bibinfo {pages} {75}
  (\bibinfo {year} {1983})}\BibitemShut {NoStop}%
\bibitem [{\citenamefont {Murphy}\ \emph {et~al.}(1994)\citenamefont {Murphy},
  \citenamefont {Eisenstein}, \citenamefont {Boebinger}, \citenamefont
  {Pfeiffer},\ and\ \citenamefont {West}}]{Murphy1994}%
  \BibitemOpen
  \bibfield  {author} {\bibinfo {author} {\bibfnamefont {S.~Q.}\ \bibnamefont
  {Murphy}}, \bibinfo {author} {\bibfnamefont {J.~P.}\ \bibnamefont
  {Eisenstein}}, \bibinfo {author} {\bibfnamefont {G.~S.}\ \bibnamefont
  {Boebinger}}, \bibinfo {author} {\bibfnamefont {L.~N.}\ \bibnamefont
  {Pfeiffer}}, \ and\ \bibinfo {author} {\bibfnamefont {K.~W.}\ \bibnamefont
  {West}},\ }\href {\doibase 10.1103/PhysRevLett.72.728} {\bibfield  {journal}
  {\bibinfo  {journal} {Phys. Rev. Lett.}\ }\textbf {\bibinfo {volume} {72}},\
  \bibinfo {pages} {728} (\bibinfo {year} {1994})}\BibitemShut {NoStop}%
\bibitem [{\citenamefont {Kellogg}\ \emph {et~al.}(2004)\citenamefont
  {Kellogg}, \citenamefont {Eisenstein}, \citenamefont {Pfeiffer},\ and\
  \citenamefont {West}}]{Kellogg2004}%
  \BibitemOpen
  \bibfield  {author} {\bibinfo {author} {\bibfnamefont {M.}~\bibnamefont
  {Kellogg}}, \bibinfo {author} {\bibfnamefont {J.~P.}\ \bibnamefont
  {Eisenstein}}, \bibinfo {author} {\bibfnamefont {L.~N.}\ \bibnamefont
  {Pfeiffer}}, \ and\ \bibinfo {author} {\bibfnamefont {K.~W.}\ \bibnamefont
  {West}},\ }\href {\doibase 10.1103/PhysRevLett.93.036801} {\bibfield
  {journal} {\bibinfo  {journal} {Phys. Rev. Lett.}\ }\textbf {\bibinfo
  {volume} {93}},\ \bibinfo {pages} {036801} (\bibinfo {year}
  {2004})}\BibitemShut {NoStop}%
\bibitem [{\citenamefont {Tutuc}\ \emph {et~al.}(2004)\citenamefont {Tutuc},
  \citenamefont {Shayegan},\ and\ \citenamefont {Huse}}]{Tutuc2004}%
  \BibitemOpen
  \bibfield  {author} {\bibinfo {author} {\bibfnamefont {E.}~\bibnamefont
  {Tutuc}}, \bibinfo {author} {\bibfnamefont {M.}~\bibnamefont {Shayegan}}, \
  and\ \bibinfo {author} {\bibfnamefont {D.~A.}\ \bibnamefont {Huse}},\ }\href
  {\doibase 10.1103/PhysRevLett.93.036802} {\bibfield  {journal} {\bibinfo
  {journal} {Phys. Rev. Lett.}\ }\textbf {\bibinfo {volume} {93}},\ \bibinfo
  {pages} {036802} (\bibinfo {year} {2004})}\BibitemShut {NoStop}%
\bibitem [{\citenamefont {Yoon}\ \emph {et~al.}(2010)\citenamefont {Yoon},
  \citenamefont {Tiemann}, \citenamefont {Schmult}, \citenamefont {Dietsche},
  \citenamefont {von Klitzing},\ and\ \citenamefont {Wegscheider}}]{Yoon2010}%
  \BibitemOpen
  \bibfield  {author} {\bibinfo {author} {\bibfnamefont {Y.}~\bibnamefont
  {Yoon}}, \bibinfo {author} {\bibfnamefont {L.}~\bibnamefont {Tiemann}},
  \bibinfo {author} {\bibfnamefont {S.}~\bibnamefont {Schmult}}, \bibinfo
  {author} {\bibfnamefont {W.}~\bibnamefont {Dietsche}}, \bibinfo {author}
  {\bibfnamefont {K.}~\bibnamefont {von Klitzing}}, \ and\ \bibinfo {author}
  {\bibfnamefont {W.}~\bibnamefont {Wegscheider}},\ }\href {\doibase
  10.1103/PhysRevLett.104.116802} {\bibfield  {journal} {\bibinfo  {journal}
  {Phys. Rev. Lett.}\ }\textbf {\bibinfo {volume} {104}},\ \bibinfo {pages}
  {116802} (\bibinfo {year} {2010})}\BibitemShut {NoStop}%
\bibitem [{\citenamefont {Spielman}(2004)}]{SpielmanPhD}%
  \BibitemOpen
  \bibfield  {author} {\bibinfo {author} {\bibfnamefont {I.~B.}\ \bibnamefont
  {Spielman}},\ }\emph {\bibinfo {title} {Evidence for the Josephson Effect in
  Quantum Hall Bilayers}},\ \href@noop {} {Ph.D. thesis},\ \bibinfo  {school}
  {California Institute of Technology} (\bibinfo {year} {2004})\BibitemShut
  {NoStop}%
\bibitem [{\citenamefont {Tiemann}\ \emph {et~al.}(2008)\citenamefont
  {Tiemann}, \citenamefont {Dietsche}, \citenamefont {Hauser},\ and\
  \citenamefont {von Klitzing}}]{TiemannNJP2008}%
  \BibitemOpen
  \bibfield  {author} {\bibinfo {author} {\bibfnamefont {L.}~\bibnamefont
  {Tiemann}}, \bibinfo {author} {\bibfnamefont {W.}~\bibnamefont {Dietsche}},
  \bibinfo {author} {\bibfnamefont {M.}~\bibnamefont {Hauser}}, \ and\ \bibinfo
  {author} {\bibfnamefont {K.}~\bibnamefont {von Klitzing}},\ }\href@noop {}
  {\bibfield  {journal} {\bibinfo  {journal} {New J. Phys.}\ }\textbf {\bibinfo
  {volume} {10}},\ \bibinfo {pages} {045018} (\bibinfo {year}
  {2008})}\BibitemShut {NoStop}%
\bibitem [{\citenamefont {Desrat}\ \emph {et~al.}(2002)\citenamefont {Desrat},
  \citenamefont {Maude}, \citenamefont {Potemski}, \citenamefont {Portal},
  \citenamefont {Wasilewski},\ and\ \citenamefont {Hill}}]{Desrat2002}%
  \BibitemOpen
  \bibfield  {author} {\bibinfo {author} {\bibfnamefont {W.}~\bibnamefont
  {Desrat}}, \bibinfo {author} {\bibfnamefont {D.~K.}\ \bibnamefont {Maude}},
  \bibinfo {author} {\bibfnamefont {M.}~\bibnamefont {Potemski}}, \bibinfo
  {author} {\bibfnamefont {J.~C.}\ \bibnamefont {Portal}}, \bibinfo {author}
  {\bibfnamefont {Z.~R.}\ \bibnamefont {Wasilewski}}, \ and\ \bibinfo {author}
  {\bibfnamefont {G.}~\bibnamefont {Hill}},\ }\href {\doibase
  10.1103/PhysRevLett.88.256807} {\bibfield  {journal} {\bibinfo  {journal}
  {Phys. Rev. Lett.}\ }\textbf {\bibinfo {volume} {88}},\ \bibinfo {pages}
  {256807} (\bibinfo {year} {2002})}\BibitemShut {NoStop}%
\bibitem [{\citenamefont {Tiemann}\ \emph {et~al.}(2012)\citenamefont
  {Tiemann}, \citenamefont {Gamez}, \citenamefont {Kumada},\ and\ \citenamefont
  {Muraki}}]{Tiemann2012}%
  \BibitemOpen
  \bibfield  {author} {\bibinfo {author} {\bibfnamefont {L.}~\bibnamefont
  {Tiemann}}, \bibinfo {author} {\bibfnamefont {G.}~\bibnamefont {Gamez}},
  \bibinfo {author} {\bibfnamefont {N.}~\bibnamefont {Kumada}}, \ and\ \bibinfo
  {author} {\bibfnamefont {K.}~\bibnamefont {Muraki}},\ }\href@noop {}
  {\bibfield  {journal} {\bibinfo  {journal} {Science}\ }\textbf {\bibinfo
  {volume} {335}},\ \bibinfo {pages} {828} (\bibinfo {year}
  {2012})}\BibitemShut {NoStop}%
\bibitem [{\citenamefont {Tsui}\ \emph {et~al.}(1982)\citenamefont {Tsui},
  \citenamefont {Stormer},\ and\ \citenamefont {Gossard}}]{Tsui82}%
  \BibitemOpen
  \bibfield  {author} {\bibinfo {author} {\bibfnamefont {D.~C.}\ \bibnamefont
  {Tsui}}, \bibinfo {author} {\bibfnamefont {H.~L.}\ \bibnamefont {Stormer}}, \
  and\ \bibinfo {author} {\bibfnamefont {A.~C.}\ \bibnamefont {Gossard}},\
  }\href {\doibase 10.1103/PhysRevLett.48.1559} {\bibfield  {journal} {\bibinfo
   {journal} {Phys. Rev. Lett.}\ }\textbf {\bibinfo {volume} {48}},\ \bibinfo
  {pages} {1559} (\bibinfo {year} {1982})}\BibitemShut {NoStop}%
\bibitem [{\citenamefont {Laughlin}(1983)}]{Laughlin83}%
  \BibitemOpen
  \bibfield  {author} {\bibinfo {author} {\bibfnamefont {R.~B.}\ \bibnamefont
  {Laughlin}},\ }\href {\doibase 10.1103/PhysRevLett.50.1395} {\bibfield
  {journal} {\bibinfo  {journal} {Phys. Rev. Lett.}\ }\textbf {\bibinfo
  {volume} {50}},\ \bibinfo {pages} {1395} (\bibinfo {year}
  {1983})}\BibitemShut {NoStop}%
\bibitem [{\citenamefont {Fertig}(1989)}]{Fertig1990}%
  \BibitemOpen
  \bibfield  {author} {\bibinfo {author} {\bibfnamefont {H.~A.}\ \bibnamefont
  {Fertig}},\ }\href {\doibase 10.1103/PhysRevB.40.1087} {\bibfield  {journal}
  {\bibinfo  {journal} {Phys. Rev. B}\ }\textbf {\bibinfo {volume} {40}},\
  \bibinfo {pages} {1087} (\bibinfo {year} {1989})}\BibitemShut {NoStop}%
\bibitem [{\citenamefont {MacDonald}\ and\ \citenamefont
  {Rezayi}(1990)}]{MacDonald1990}%
  \BibitemOpen
  \bibfield  {author} {\bibinfo {author} {\bibfnamefont {A.~H.}\ \bibnamefont
  {MacDonald}}\ and\ \bibinfo {author} {\bibfnamefont {E.~H.}\ \bibnamefont
  {Rezayi}},\ }\href {\doibase 10.1103/PhysRevB.42.3224} {\bibfield  {journal}
  {\bibinfo  {journal} {Phys. Rev. B}\ }\textbf {\bibinfo {volume} {42}},\
  \bibinfo {pages} {3224} (\bibinfo {year} {1990})}\BibitemShut {NoStop}%
\bibitem [{\citenamefont {Moon}\ \emph {et~al.}(1995)\citenamefont {Moon},
  \citenamefont {Mori}, \citenamefont {Yang}, \citenamefont {Girvin},
  \citenamefont {MacDonald}, \citenamefont {Zheng}, \citenamefont {Yoshioka},\
  and\ \citenamefont {Zhang}}]{Moon1995}%
  \BibitemOpen
  \bibfield  {author} {\bibinfo {author} {\bibfnamefont {K.}~\bibnamefont
  {Moon}}, \bibinfo {author} {\bibfnamefont {H.}~\bibnamefont {Mori}}, \bibinfo
  {author} {\bibfnamefont {K.}~\bibnamefont {Yang}}, \bibinfo {author}
  {\bibfnamefont {S.~M.}\ \bibnamefont {Girvin}}, \bibinfo {author}
  {\bibfnamefont {A.~H.}\ \bibnamefont {MacDonald}}, \bibinfo {author}
  {\bibfnamefont {L.}~\bibnamefont {Zheng}}, \bibinfo {author} {\bibfnamefont
  {D.}~\bibnamefont {Yoshioka}}, \ and\ \bibinfo {author} {\bibfnamefont
  {S.-C.}\ \bibnamefont {Zhang}},\ }\href {\doibase 10.1103/PhysRevB.51.5138}
  {\bibfield  {journal} {\bibinfo  {journal} {Phys. Rev. B}\ }\textbf {\bibinfo
  {volume} {51}},\ \bibinfo {pages} {5138} (\bibinfo {year}
  {1995})}\BibitemShut {NoStop}%
\bibitem [{\citenamefont {Ezawa}(2013)}]{Ezawa2013}%
  \BibitemOpen
  \bibfield  {author} {\bibinfo {author} {\bibfnamefont {F.~Z.}\ \bibnamefont
  {Ezawa}},\ }\href@noop {} {\emph {\bibinfo {title} {{Quantum Hall Effects -
  Recent Theoretical and Experimental Developments}}}}\ (\bibinfo  {publisher}
  {World Scientific},\ \bibinfo {year} {2013})\BibitemShut {NoStop}%
\bibitem [{\citenamefont {Giudici}\ \emph {et~al.}(2010)\citenamefont
  {Giudici}, \citenamefont {Muraki}, \citenamefont {Kumada},\ and\
  \citenamefont {Fujisawa}}]{Giudici2010}%
  \BibitemOpen
  \bibfield  {author} {\bibinfo {author} {\bibfnamefont {P.}~\bibnamefont
  {Giudici}}, \bibinfo {author} {\bibfnamefont {K.}~\bibnamefont {Muraki}},
  \bibinfo {author} {\bibfnamefont {N.}~\bibnamefont {Kumada}}, \ and\ \bibinfo
  {author} {\bibfnamefont {T.}~\bibnamefont {Fujisawa}},\ }\href {\doibase
  10.1103/PhysRevLett.104.056802} {\bibfield  {journal} {\bibinfo  {journal}
  {Phys. Rev. Lett.}\ }\textbf {\bibinfo {volume} {104}},\ \bibinfo {pages}
  {056802} (\bibinfo {year} {2010})}\BibitemShut {NoStop}%
\bibitem [{\citenamefont {Kumada}\ \emph {et~al.}(2005)\citenamefont {Kumada},
  \citenamefont {Muraki}, \citenamefont {Hashimoto},\ and\ \citenamefont
  {Y}}]{Kumada2005}%
  \BibitemOpen
  \bibfield  {author} {\bibinfo {author} {\bibfnamefont {N.}~\bibnamefont
  {Kumada}}, \bibinfo {author} {\bibfnamefont {K.}~\bibnamefont {Muraki}},
  \bibinfo {author} {\bibfnamefont {K.}~\bibnamefont {Hashimoto}}, \ and\
  \bibinfo {author} {\bibfnamefont {Y.}~\bibnamefont {Hirayama}},\ }\href {\doibase
  10.1103/PhysRevLett.94.096802} {\bibfield  {journal} {\bibinfo  {journal}
  {Phys. Rev. Lett.}\ }\textbf {\bibinfo {volume} {94}},\ \bibinfo {pages}
  {096802} (\bibinfo {year} {2005})}\BibitemShut {NoStop}%
\bibitem [{\citenamefont {Finck}\ \emph {et~al.}(2010)\citenamefont {Finck},
  \citenamefont {Eisenstein}, \citenamefont {Pfeiffer},\ and\ \citenamefont
  {West}}]{Finck2010}%
  \BibitemOpen
  \bibfield  {author} {\bibinfo {author} {\bibfnamefont {A.~D.~K.}\
  \bibnamefont {Finck}}, \bibinfo {author} {\bibfnamefont {J.~P.}\ \bibnamefont
  {Eisenstein}}, \bibinfo {author} {\bibfnamefont {L.~N.}\ \bibnamefont
  {Pfeiffer}}, \ and\ \bibinfo {author} {\bibfnamefont {K.~W.}\ \bibnamefont
  {West}},\ }\href {\doibase 10.1103/PhysRevLett.104.016801} {\bibfield
  {journal} {\bibinfo  {journal} {Phys. Rev. Lett.}\ }\textbf {\bibinfo
  {volume} {104}},\ \bibinfo {pages} {016801} (\bibinfo {year}
  {2010})}\BibitemShut {NoStop}%
\bibitem [{\citenamefont {Spielman}\ \emph {et~al.}(2004)\citenamefont
  {Spielman}, \citenamefont {Kellogg}, \citenamefont {Eisenstein},
  \citenamefont {Pfeiffer},\ and\ \citenamefont {West}}]{Spielman2004}%
  \BibitemOpen
  \bibfield  {author} {\bibinfo {author} {\bibfnamefont {I.~B.}\ \bibnamefont
  {Spielman}}, \bibinfo {author} {\bibfnamefont {M.}~\bibnamefont {Kellogg}},
  \bibinfo {author} {\bibfnamefont {J.~P.}\ \bibnamefont {Eisenstein}},
  \bibinfo {author} {\bibfnamefont {L.~N.}\ \bibnamefont {Pfeiffer}}, \ and\
  \bibinfo {author} {\bibfnamefont {K.~W.}\ \bibnamefont {West}},\ }\href
  {\doibase 10.1103/PhysRevB.70.081303} {\bibfield  {journal} {\bibinfo
  {journal} {Phys. Rev. B}\ }\textbf {\bibinfo {volume} {70}},\ \bibinfo
  {pages} {081303} (\bibinfo {year} {2004})}\BibitemShut {NoStop}%
\bibitem [{\citenamefont {Wiersma}\ \emph {et~al.}(2004)\citenamefont
  {Wiersma}, \citenamefont {Lok}, \citenamefont {Kraus}, \citenamefont
  {Dietsche}, \citenamefont {von Klitzing}, \citenamefont {Schuh},
  \citenamefont {Bichler}, \citenamefont {Tranitz},\ and\ \citenamefont
  {Wegscheider}}]{Wiersma2004}%
  \BibitemOpen
  \bibfield  {author} {\bibinfo {author} {\bibfnamefont {R.~D.}\ \bibnamefont
  {Wiersma}}, \bibinfo {author} {\bibfnamefont {J.~G.~S.}\ \bibnamefont {Lok}},
  \bibinfo {author} {\bibfnamefont {S.}~\bibnamefont {Kraus}}, \bibinfo
  {author} {\bibfnamefont {W.}~\bibnamefont {Dietsche}}, \bibinfo {author}
  {\bibfnamefont {K.}~\bibnamefont {von Klitzing}}, \bibinfo {author}
  {\bibfnamefont {D.}~\bibnamefont {Schuh}}, \bibinfo {author} {\bibfnamefont
  {M.}~\bibnamefont {Bichler}}, \bibinfo {author} {\bibfnamefont {H.-P.}\
  \bibnamefont {Tranitz}}, \ and\ \bibinfo {author} {\bibfnamefont
  {W.}~\bibnamefont {Wegscheider}},\ }\href {\doibase
  10.1103/PhysRevLett.93.266805} {\bibfield  {journal} {\bibinfo  {journal}
  {Phys. Rev. Lett.}\ }\textbf {\bibinfo {volume} {93}},\ \bibinfo {pages}
  {266805} (\bibinfo {year} {2004})}\BibitemShut {NoStop}%
\bibitem [{\citenamefont {Champagne}\ \emph {et~al.}(2008)\citenamefont
  {Champagne}, \citenamefont {Finck}, \citenamefont {Eisenstein}, \citenamefont
  {Pfeiffer},\ and\ \citenamefont {West}}]{Champagne2008}%
  \BibitemOpen
  \bibfield  {author} {\bibinfo {author} {\bibfnamefont {A.~R.}\ \bibnamefont
  {Champagne}}, \bibinfo {author} {\bibfnamefont {A.~D.~K.}\ \bibnamefont
  {Finck}}, \bibinfo {author} {\bibfnamefont {J.~P.}\ \bibnamefont
  {Eisenstein}}, \bibinfo {author} {\bibfnamefont {L.~N.}\ \bibnamefont
  {Pfeiffer}}, \ and\ \bibinfo {author} {\bibfnamefont {K.~W.}\ \bibnamefont
  {West}},\ }\href {\doibase 10.1103/PhysRevB.78.205310} {\bibfield  {journal}
  {\bibinfo  {journal} {Phys. Rev. B}\ }\textbf {\bibinfo {volume} {78}},\
  \bibinfo {pages} {205310} (\bibinfo {year} {2008})}\BibitemShut {NoStop}%
\bibitem [{\citenamefont {Takase}\ and\ \citenamefont
  {Muraki}(2011)}]{Takase2011}%
  \BibitemOpen
  \bibfield  {author} {\bibinfo {author} {\bibfnamefont {K.}~\bibnamefont
  {Takase}}\ and\ \bibinfo {author} {\bibfnamefont {K.}~\bibnamefont
  {Muraki}},\ }\href {\doibase 10.1088/1742-6596/334/1/012025} {\bibfield
  {journal} {\bibinfo  {journal} {Journal of Physics: Conference Series}\
  }\textbf {\bibinfo {volume} {334}},\ \bibinfo {pages} {012025} (\bibinfo
  {year} {2011})}\BibitemShut {NoStop}%
\bibitem [{\citenamefont {Zhang}\ \emph {et~al.}(2013)\citenamefont {Zhang},
  \citenamefont {Schmult}, \citenamefont {Venkatachalam}, \citenamefont
  {Dietsche}, \citenamefont {Yacoby}, \citenamefont {von Klitzing},\ and\
  \citenamefont {Smet}}]{Ding2013}%
  \BibitemOpen
  \bibfield  {author} {\bibinfo {author} {\bibfnamefont {D.}~\bibnamefont
  {Zhang}}, \bibinfo {author} {\bibfnamefont {S.}~\bibnamefont {Schmult}},
  \bibinfo {author} {\bibfnamefont {V.}~\bibnamefont {Venkatachalam}}, \bibinfo
  {author} {\bibfnamefont {W.}~\bibnamefont {Dietsche}}, \bibinfo {author}
  {\bibfnamefont {A.}~\bibnamefont {Yacoby}}, \bibinfo {author} {\bibfnamefont
  {K.}~\bibnamefont {von Klitzing}}, \ and\ \bibinfo {author} {\bibfnamefont
  {J.~H.}\ \bibnamefont {Smet}},\ }\href {\doibase 10.1103/PhysRevB.87.205304}
  {\bibfield  {journal} {\bibinfo  {journal} {Phys. Rev. B}\ }\textbf {\bibinfo
  {volume} {87}},\ \bibinfo {pages} {205304} (\bibinfo {year}
  {2013})}\BibitemShut {NoStop}%
\bibitem [{\citenamefont {Eisenstein}\ \emph {et~al.}(1990)\citenamefont
  {Eisenstein}, \citenamefont {Pfeiffer},\ and\ \citenamefont
  {West}}]{Eisenstein1990b}%
  \BibitemOpen
  \bibfield  {author} {\bibinfo {author} {\bibfnamefont {J.~P.}\ \bibnamefont
  {Eisenstein}}, \bibinfo {author} {\bibfnamefont {L.~N.}\ \bibnamefont
  {Pfeiffer}}, \ and\ \bibinfo {author} {\bibfnamefont {K.~W.}\ \bibnamefont
  {West}},\ }\href@noop {} {\bibfield  {journal} {\bibinfo  {journal} {Applied
  Physics Letters}\ }\textbf {\bibinfo {volume} {57}},\ \bibinfo {pages} {2324}
  (\bibinfo {year} {1990})}\BibitemShut {NoStop}%
\bibitem [{\citenamefont {Rubel}\ \emph {et~al.}(1997)\citenamefont {Rubel},
  \citenamefont {Fischer}, \citenamefont {Dietsche}, \citenamefont {von
  Klitzing},\ and\ \citenamefont {Eberl}}]{Rubel1997}%
  \BibitemOpen
  \bibfield  {author} {\bibinfo {author} {\bibfnamefont {H.}~\bibnamefont
  {Rubel}}, \bibinfo {author} {\bibfnamefont {A.}~\bibnamefont {Fischer}},
  \bibinfo {author} {\bibfnamefont {W.}~\bibnamefont {Dietsche}}, \bibinfo
  {author} {\bibfnamefont {K.}~\bibnamefont {von Klitzing}}, \ and\ \bibinfo
  {author} {\bibfnamefont {K.}~\bibnamefont {Eberl}},\ }\href {\doibase
  10.1103/PhysRevLett.78.1763} {\bibfield  {journal} {\bibinfo  {journal}
  {Phys. Rev. Lett.}\ }\textbf {\bibinfo {volume} {78}},\ \bibinfo {pages}
  {1763} (\bibinfo {year} {1997})}\BibitemShut {NoStop}%
\bibitem [{Note1()}]{Note1}%
  \BibitemOpen
  \bibinfo {note} {Each layer acts as a gate for the adjacent layer due to
  ''compressibility'' [Phys. Rev. B 50, 1760 (1994)]. When we create a density
  imbalance $(n_1 \pm \Delta _n)$ and $(n_2 \mp \Delta n)$, this weak mutual
  gate effect also changes. For the maximal imbalance, the density would
  deviate by a few percent from the original (matched) value if we used the
  same gate voltages. To ensure that we operate at the same $\nu $ under
  imbalance, we have compensated this effect by carefully adjusting top and
  back gate voltages}\BibitemShut {NoStop}%
\bibitem [{\citenamefont {Li}\ and\ \citenamefont {Smet}(2008)}]{LiBook}%
  \BibitemOpen
  \bibfield  {author} {\bibinfo {author} {\bibfnamefont {Y.~Q.}\ \bibnamefont
  {Li}}\ and\ \bibinfo {author} {\bibfnamefont {J.~H.}\ \bibnamefont {Smet}},\
  }in\ \href@noop {} {\emph {\bibinfo {booktitle} {Spin Physics in
  Semiconductors}}},\ \bibinfo {editor} {edited by\ \bibinfo {editor}
  {\bibfnamefont {M.~I.~D.}\ \bibnamefont {(Ed.)}}}\ (\bibinfo  {publisher}
  {Springer},\ \bibinfo {year} {2008})\BibitemShut {NoStop}%
\bibitem [{\citenamefont {Stern}\ \emph {et~al.}(2012)\citenamefont {Stern},
  \citenamefont {Piot}, \citenamefont {Vardi}, \citenamefont {Umansky},
  \citenamefont {Plochocka}, \citenamefont {Maude},\ and\ \citenamefont
  {Bar-Joseph}}]{Stern2012}%
  \BibitemOpen
  \bibfield  {author} {\bibinfo {author} {\bibfnamefont {M.}~\bibnamefont
  {Stern}}, \bibinfo {author} {\bibfnamefont {B.~A.}\ \bibnamefont {Piot}},
  \bibinfo {author} {\bibfnamefont {Y.}~\bibnamefont {Vardi}}, \bibinfo
  {author} {\bibfnamefont {V.}~\bibnamefont {Umansky}}, \bibinfo {author}
  {\bibfnamefont {P.}~\bibnamefont {Plochocka}}, \bibinfo {author}
  {\bibfnamefont {D.~K.}\ \bibnamefont {Maude}}, \ and\ \bibinfo {author}
  {\bibfnamefont {I.}~\bibnamefont {Bar-Joseph}},\ }\href {\doibase
  10.1103/PhysRevLett.108.066810} {\bibfield  {journal} {\bibinfo  {journal}
  {Phys. Rev. Lett.}\ }\textbf {\bibinfo {volume} {108}},\ \bibinfo {pages}
  {066810} (\bibinfo {year} {2012})}\BibitemShut {NoStop}%
\bibitem [{\citenamefont {Friess}\ \emph {et~al.}(2014)\citenamefont {Friess},
  \citenamefont {Umansky}, \citenamefont {Tiemann}, \citenamefont {von
  Klitzing},\ and\ \citenamefont {Smet}}]{Friess2014}%
  \BibitemOpen
  \bibfield  {author} {\bibinfo {author} {\bibfnamefont {B.}~\bibnamefont
  {Friess}}, \bibinfo {author} {\bibfnamefont {V.}~\bibnamefont {Umansky}},
  \bibinfo {author} {\bibfnamefont {L.}~\bibnamefont {Tiemann}}, \bibinfo
  {author} {\bibfnamefont {K.}~\bibnamefont {von Klitzing}}, \ and\ \bibinfo
  {author} {\bibfnamefont {J.~H.}\ \bibnamefont {Smet}},\ }\href {\doibase
  10.1103/PhysRevLett.113.076803} {\bibfield  {journal} {\bibinfo  {journal}
  {Phys. Rev. Lett.}\ }\textbf {\bibinfo {volume} {113}},\ \bibinfo {pages}
  {076803} (\bibinfo {year} {2014})}\BibitemShut {NoStop}%
\bibitem [{\citenamefont {Freytag}\ \emph {et~al.}(2001)\citenamefont
  {Freytag}, \citenamefont {Tokunaga}, \citenamefont {Horvatic}, \citenamefont
  {Berthier}, \citenamefont {Shayegan},\ and\ \citenamefont
  {Levy}}]{Freytag2001}%
  \BibitemOpen
  \bibfield  {author} {\bibinfo {author} {\bibfnamefont {N.}~\bibnamefont
  {Freytag}}, \bibinfo {author} {\bibfnamefont {Y.}~\bibnamefont {Tokunaga}},
  \bibinfo {author} {\bibfnamefont {M.}~\bibnamefont {Horvatic}}, \bibinfo
  {author} {\bibfnamefont {C.}~\bibnamefont {Berthier}}, \bibinfo {author}
  {\bibfnamefont {M.}~\bibnamefont {Shayegan}}, \ and\ \bibinfo {author}
  {\bibfnamefont {L.~P.}\ \bibnamefont {Levy}},\ }\href {\doibase
  10.1103/PhysRevLett.87.136801} {\bibfield  {journal} {\bibinfo  {journal}
  {Phys. Rev. Lett.}\ }\textbf {\bibinfo {volume} {87}},\ \bibinfo {pages}
  {136801} (\bibinfo {year} {2001})}\BibitemShut {NoStop}%
\bibitem [{\citenamefont {Hayakawa}\ \emph {et~al.}(2013)\citenamefont
  {Hayakawa}, \citenamefont {Muraki},\ and\ \citenamefont {Yusa}}]{Yusa2013}%
  \BibitemOpen
  \bibfield  {author} {\bibinfo {author} {\bibfnamefont {J.}~\bibnamefont
  {Hayakawa}}, \bibinfo {author} {\bibfnamefont {K.}~\bibnamefont {Muraki}}, \
  and\ \bibinfo {author} {\bibfnamefont {G.}~\bibnamefont {Yusa}},\ }\href
  {\doibase 10.1038/nnano.2012.209} {\bibfield  {journal} {\bibinfo  {journal}
  {Nature Nanotechnology}\ }\textbf {\bibinfo {volume} {8}},\ \bibinfo {pages}
  {31} (\bibinfo {year} {2013})}\BibitemShut {NoStop}%
\bibitem [{\citenamefont {Chakraborty}(1990)}]{Chakraborty1990}%
  \BibitemOpen
  \bibfield  {author} {\bibinfo {author} {\bibfnamefont {T.}~\bibnamefont
  {Chakraborty}},\ }\href@noop {} {\bibfield  {journal} {\bibinfo  {journal}
  {Surface Science}\ }\textbf {\bibinfo {volume} {229}},\ \bibinfo {pages} {16}
  (\bibinfo {year} {1990})}\BibitemShut {NoStop}%
\bibitem [{\citenamefont {Hoeppel}(2004)}]{HoeppelPhD}%
  \BibitemOpen
  \bibfield  {author} {\bibinfo {author} {\bibfnamefont {L.~W.}\ \bibnamefont
  {Hoeppel}},\ }\emph {\bibinfo {title} {{Phase transitions in single layer and
  bilayer quantum Hall ferromagnets}}},\ \href@noop {} {Ph.D. thesis},\
  \bibinfo  {school} {University of Stuttgart} (\bibinfo {year}
  {2004})\BibitemShut {NoStop}%
\bibitem [{\citenamefont {Zhang}\ and\ \citenamefont
  {Sarma}(1986)}]{DasSarma1986}%
  \BibitemOpen
  \bibfield  {author} {\bibinfo {author} {\bibfnamefont {F.~C.}\ \bibnamefont
  {Zhang}}\ and\ \bibinfo {author} {\bibfnamefont {S.}\ \bibnamefont
  {DasSarma}},\ }\href {\doibase 10.1103/PhysRevB.33.2903} {\bibfield  {journal}
  {\bibinfo  {journal} {Phys. Rev. B}\ }\textbf {\bibinfo {volume} {33}},\
  \bibinfo {pages} {2903} (\bibinfo {year} {1986})}\BibitemShut {NoStop}%
\bibitem [{Note2()}]{Note2}%
  \BibitemOpen
  \bibinfo {note} {We have scrutinized the temperature dependence, but due to
  the small Knight shift we found no significant/discernable change in $K_s$
  for $(\nu _1,\nu _2) = ({\begingroup 1\endgroup \over 3},{\begingroup
  2\endgroup \over 3})$ up to ca. 300 mK at which the NMR signal became too
  small.}\BibitemShut {Stop}%
\bibitem [{\citenamefont {Spielman}\ \emph {et~al.}(2005)\citenamefont
  {Spielman}, \citenamefont {Tracy}, \citenamefont {Eisenstein}, \citenamefont
  {Pfeiffer},\ and\ \citenamefont {West}}]{Spielman2005}%
  \BibitemOpen
  \bibfield  {author} {\bibinfo {author} {\bibfnamefont {I.~B.}\ \bibnamefont
  {Spielman}}, \bibinfo {author} {\bibfnamefont {L.~A.}\ \bibnamefont {Tracy}},
  \bibinfo {author} {\bibfnamefont {J.~P.}\ \bibnamefont {Eisenstein}},
  \bibinfo {author} {\bibfnamefont {L.~N.}\ \bibnamefont {Pfeiffer}}, \ and\
  \bibinfo {author} {\bibfnamefont {K.~W.}\ \bibnamefont {West}},\ }\href
  {\doibase 10.1103/PhysRevLett.94.076803} {\bibfield  {journal} {\bibinfo
  {journal} {Phys. Rev. Lett.}\ }\textbf {\bibinfo {volume} {94}},\ \bibinfo
  {pages} {076803} (\bibinfo {year} {2005})}\BibitemShut {NoStop}%
\bibitem [{\citenamefont {Tracy}(2007)}]{TracyPhD}%
  \BibitemOpen
  \bibfield  {author} {\bibinfo {author} {\bibfnamefont {L.~A.}\ \bibnamefont
  {Tracy}},\ }\emph {\bibinfo {title} {{Studies of Two Dimensional Electron
  Systems via Surface Acoustic Waves and Nuclear Magnetic Resonance
  Techniques}}},\ \href@noop {} {Ph.D. thesis},\ \bibinfo  {school} {California
  Institute of Technology} (\bibinfo {year} {2007})\BibitemShut {NoStop}%
\bibitem [{\citenamefont {Kukushkin}\ \emph {et~al.}(1999)\citenamefont
  {Kukushkin}, \citenamefont {v.~Klitzing},\ and\ \citenamefont
  {Eberl}}]{Igor1999}%
  \BibitemOpen
  \bibfield  {author} {\bibinfo {author} {\bibfnamefont {I.~V.}\ \bibnamefont
  {Kukushkin}}, \bibinfo {author} {\bibfnamefont {K.}~\bibnamefont
  {v.~Klitzing}}, \ and\ \bibinfo {author} {\bibfnamefont {K.}~\bibnamefont
  {Eberl}},\ }\href {\doibase 10.1103/PhysRevLett.82.3665} {\bibfield
  {journal} {\bibinfo  {journal} {Phys. Rev. Lett.}\ }\textbf {\bibinfo
  {volume} {82}},\ \bibinfo {pages} {3665} (\bibinfo {year}
  {1999})}\BibitemShut {NoStop}%
\end{thebibliography}%

\end{document}